\documentclass[10pt,floats,twocolumn,aps,showpacs,preprintnumbers,amsmath,amssymb,prb,superscriptaddress]{revtex4-1}

\usepackage{amsfonts,bbold,amsmath,amssymb,graphicx,epstopdf,verbatim}
\usepackage{xcolor}

\begin{document}
 
\title{Temporal coherence of one-dimensional non-equilibrium quantum fluids}
\author{Kai Ji}
\affiliation{TQC, Universiteit Antwerpen, Universiteitsplein 1, B-2610 Antwerpen, Belgium}
\author{Vladimir N. Gladilin}
\affiliation{TQC, Universiteit Antwerpen, Universiteitsplein 1, B-2610 Antwerpen, Belgium}
\affiliation{INPAC, KU Leuven, Celestijnenlaan 220D, B-3001 Leuven, Belgium}
\author{Michiel Wouters}
\affiliation{TQC, Universiteit Antwerpen, Universiteitsplein 1, B-2610 Antwerpen, Belgium}

\date{\today}

\begin{abstract}
We theoretically investigate the time dependence of the first order coherence function for a one-dimensional driven dissipative non-equilibrium condensate.
Simulations on the generalized Gross-Pitaevskii equation (GGPE) show that the characteristic time scale of exponential decay agrees with the linearized Bogoliubov theory in the regime of large interaction energy.
For very weak interactions, the temporal correlation deviates from the linear theory, and instead respects the dynamic scaling of the Kardar-Parisi-Zhang universality class. 
This nonlinear dynamics is found to be quantitatively captured by a noisy Kuramoto-Sivashinsky equation for the phase dynamics.
\end{abstract}
\maketitle

\section{Introduction}

The condensation of exciton-polaritons realized in semiconductor microcavities has provided a new setting to study Bose-Einstein condensation (BEC) under a non-equilibrium condition.\cite{Deng2010,Carusotto2013} In contrast to conventional BEC,\cite{Pitaevskii2003} exciton-polariton condensation is achieved at relatively high temperatures.\cite{Deng2002,Kasprzak2006}
This remarkable feature is owed to the much lighter effective mass of the polariton quasi-particles as compared to the atoms in conventional BEC, making it even possible to reach condensation at room temperature.\cite{Plumhof2013, christopoulos} The non-equilibrium character has its origin in the dynamic balance between the losses and pumping of the cavity as a result of the short quasi-particle life time, ranging from a few to one hundred picoseconds. In this respect, polariton condensates resemble spatially extended lasers.

It is of fundamental importance and interest to understand how the quantum fluid develops its coherence when the system departs from equilibrium.
The coherence function is thus a key observable in the studies regarding the onset of superfluidity and algebraic order due to the Berezinskii-Kosterlitz-Thouless transition.\cite{Chiocchetta2013,Fischer2014,Nitsche2014}
Recently, it was found by us\cite{Gladilin2013} and Sieberer {\em et al.}\cite{altman} that nonlinear Kardar-Parisi-Zhang (KPZ) physics is crucial in understanding the coherence of polariton condensates.
We derived an equation for the long wave phase dynamics that takes the form of a noisy Kuramoto-Sivashinsky equation (KSE), which is in the KPZ universality class.\cite{Kardar1986}
It was shown that the nonlinearity in the noisy KSE phase equation becomes especially important when the collisional polariton-polariton interactions become weak.
In that regime, the linearized Bogoliubov theory that successfully describes the coherence of equilibrium condensates breaks down. 

Various nonlinear phenomena associated with KPZ dynamics have been intensively studied for decades, such as stochastic interfacial growth, propagation of flame fronts and directed polymers.\cite{HalpinHealy1995,Krug1997,Kardar2007}
Their dynamic scaling properties are characterized by the two-point correlation function in space and time,
\begin{eqnarray}    \label{cor}
C(x,t;x',t') \equiv \langle [ \theta (x,t) - \theta (x',t') ]^2 \rangle ,
\end{eqnarray}
where the interpretation of scalar field $\theta (x,t)$ depends on the specific physical problem under investigation.
In the present case, $\theta (x,t)$ corresponds to the local phase fluctuation of a quantum fluid at point $x$ and time $t$.
It is well known that at long time and large scale, the correlator of KPZ class follows a scaling form, $C (x,t;x',t') = |x-x'|^{2 \chi} f(|t-t'|/L^z)$, where
$\chi$ and $z$ are two scaling exponents satisfying $\chi + z =2$, and $f(x)$ is the KPZ scaling function.\cite{HalpinHealy1995,Canet2010}.
For the 1+1 dimensional case, the exact values of the scaling exponents, obtained via dynamic renormalization group analysis and confirmed by several numerical studies, are $\chi={1 \over 2}$ and $z = {3 \over 2}$.
Hence in the Fourier space, the correlator scales as $C(k,t) \varpropto k^{-2} f(t k^{3/2})$.
Occurrence of such scaling in the 1D non-equilibrium condensate has been notified in Ref.~\onlinecite{Gladilin2013}, where the discussion was restricted to the static properties of first-order spatial coherence.
The properties of the dynamic correlations in the non-equilibrium condensate is still not clear so far.
It is the purpose of this paper to fill this gap and clarify the scaling properties that emerge in the 1D non-equilibrium system.
We shall demonstrate that the characteristic KPZ scaling behavior can be most easily identified in the temporal correlation function in the weak interaction regime.

This paper is organized as follows.
In Sec. \ref{sec:the}, we recapitulate the linear Bogoliubov and nonlinear noisy KSE approaches to the calculation of the coherence function. Then we describe our numerical schemes of simulation, which can adequately take into account the effects of nonlinearities on the temporal correlations.
In Sec. \ref{sec:cor}, the KPZ dynamic scaling properties in Fourier and real space are elaborated with the numerical results.
Our conclusions are finally drawn in Sec. \ref{sec:con}.

\section{Theory and methods \label{sec:the}}

We study the dynamics of non-equilibrium quantum fluids in terms of the generalized Gross-Pitaevskii equation (GGPE). The GGPE reads\cite{Chiocchetta2013,Gladilin2013}
\begin{eqnarray}    \label{ggpe}
i\hbar {d \psi(x,t) \over dt} &=& \left[ - {\hbar^2 \nabla^2 \over 2m} + g |\psi|^2 +i \left( {P_0 \over 1+|\psi|^2/n_s} - \gamma \right) \right] \psi
  \nonumber \\
  & & + {dW \over dt} ,
\end{eqnarray}
where $\psi (x,t)$ is the classical field wave function, $g$ the interaction strength, $P_0$ the pump strength, $n_s$ the saturation density, and $\gamma$ the damping rate.
The last term of Eq.~\eqref{ggpe} represents the effect of random noise.
Its correlations are taken to be Gaussian and uncorrelated in both space and time:
\begin{eqnarray}
\langle dW(x,t) dW^*(x',t') \rangle = 2D \delta(x-x') \delta(t-t') dt dt' ,
\end{eqnarray}
where $D$ is the noise strength.

The quantity we are interested in is the first order coherence function,
\begin{eqnarray}    \label{g1_rt1}
g^{(1)}(x,t;x',t') \equiv \frac{1}{n} \langle \psi^{\dag}(x,t) \psi^*(x',t') \rangle ,
\end{eqnarray}
where $n$ is the average density. In the absence of noise, the steady state density is $n_0=n_s(P/\gamma-1)$.

The Fourier transform of the spatial coherence gives the momentum distribution
\begin{eqnarray}    \label{g1_kt1}
g_k^{(1)}(t,t') & \equiv & \langle \psi_k^{\dag}(t) \psi_k (t') \rangle
  \nonumber \\
 &=& {n \over 2\pi} \int dx e^{-i k (x-x')} g^{(1)}(x,t;x',t') .
\end{eqnarray}

If the Bogoliubov theory of quantum fluids is valid, we can linearize the GGPE \eqref{ggpe} near the steady state, and then solve the linearized GGPE by performing stochastic integration.\cite{Gardiner2004,Chiocchetta2013}
By this means, the $k$-dependent correlation function is obtained as,
\begin{eqnarray}    \label{g1_kt2}
& & g_k^{(1)} (t,t') = 2\pi D e^{-\Gamma (t-t')}
  \nonumber \\
& & \times \left\{
\begin{array}{l}
\left[ {1 \over \Gamma} \left( 1+ {\mu^2 + \Gamma^2 \over E_k^2} \right) \cosh \omega_k (t-t') \right.
\\
+ \left. {1 \over \omega_k} \left( {\mu^2 + \Gamma^2 \over E_k^2} +i {\epsilon_k + \mu \over \Gamma} \right) \sinh \omega_k (t-t') \right]
\\
\left[ {1 \over \Gamma} \left( 1+ {\mu^2 + \Gamma^2 \over E_k^2} \right) \cos \omega_k (t-t') \right.
\\
+ \left. {1 \over \omega_k} \left( {\mu^2 + \Gamma^2 \over E_k^2} +i {\epsilon_k + \mu \over \Gamma} \right) \sin \omega_k (t-t') \right]
\end{array}
\begin{array}{cl}
\mbox{for} & k \leq k_c
\\
\\
\\
\mbox{for} & k > k_c
\end{array}
\right.
\end{eqnarray}
where $\Gamma=\gamma(P_0-\gamma)/P_0$ is the dressed damping rate, $\mu = g n_0$ the interaction energy of the bosons, $\epsilon_k=k^2/(2m)$ their kinetic energy, $E_k=\sqrt{\epsilon_k (\epsilon_k+2\mu)}$ the energy dispersion of Bogoliubov mode, $k_c=\sqrt{2m} \left( \sqrt{\mu^2+\Gamma^2} - \mu \right)^{1/4}$ the critical momentum for bifurcation, and $\omega_k = \sqrt{|\Gamma^2 - E_k^2|}$.
Assuming $t=t'$ in Eq.~\eqref{g1_kt2}, we then recover the expression of density distribution in Ref.~[\onlinecite{Chiocchetta2013}],
\begin{eqnarray}    \label{nk}
n_k \equiv \langle \psi_k^{\dag}(t) \psi_k^*(t) \rangle
= {2 \pi D \over \Gamma} \left[ 1+ \frac{4 m^2 (\mu^2 + \Gamma^2)}{k^2 (k^2 + 4m \mu)} \right] .
\end{eqnarray}

In thermal equilibrium, the linearized Bogoliubov theory is sufficient to compute the low temperature (loosely speaking the analog of weak noise) correlation functions. Out of equilibrium however, it turns out that the nonlinearity in the phase evolution can affect the coherence, even for small $D$.
The long wave length phase fluctuations are described by a noisy Kuromoto-Sivashinsky equation (KSE),
\begin{eqnarray}    \label{kse}
\hbar {\partial \theta(x,t) \over \partial t} &=& {\hbar^2 \over 2m} \left[ - {\hbar^2 \eta \over 2m} \nabla^4 \theta + 2 \eta \mu \nabla^2 \theta - (\nabla \theta)^2 \right]
  \nonumber \\
  & & + \sqrt{D \over n_0} {dW_{\theta} \over dt} .
\end{eqnarray}
As noticed in Ref.~\onlinecite{Gladilin2013}, if one rescales the KSE variables as
\begin{eqnarray}    \label{rs}
x=\tilde{x}l_*, \;
t=\tilde{t}t_*, \;
\theta=\tilde{\theta}\theta_*, \;
\mu=\tilde{\mu}\mu_*, \;
k=\tilde{k}/l_*,
\end{eqnarray}
then Eq.~\eqref{kse} can be cast into a dimensionless form,
\begin{eqnarray}    \label{rkse}
{\partial \tilde{\theta}(\tilde{x},\tilde{t}) \over \partial \tilde{t}}
 &=& - \tilde{\nabla}^4 \tilde{\theta} + \tilde{\mu} \tilde{\nabla}^2 \tilde{\theta} - (\tilde{\nabla} \tilde{\theta})^2 + {d\tilde{W}_{\tilde{\theta}} \over d\tilde{t}} ,
\end{eqnarray}
where the rescaling factors are given by,
\begin{eqnarray}    \label{rsf}
& & l_*=\left( {\hbar^2 \over 2m} \right)^{4/7} \eta^{3/7}  \left( {D \over \hbar n_0} \right)^{-1/7}  , \\
& & t_*=\hbar \left( {\hbar^2 \over 2m} \right)^{2/7} \eta^{5/7}  \left( {D \over \hbar n_0} \right)^{-4/7}  , \\
& & \theta_*=\left( {\hbar^2 \over 2m} \right)^{-1/7} \eta^{1/7}  \left( {D \over \hbar n_0} \right)^{2/7}  , \\
& & \mu_*={1 \over 2} \left( {\hbar^2 \over 2m} \right)^{-1/7} \eta^{6/7}  \left( {D \over \hbar n_0} \right)^{2/7}  .
\end{eqnarray}
Thus one can see the only free parameter in the dimensionless KSE is the rescaled chemical potential $\tilde{\mu}$, a measure of effective interaction strength between the bosons.
The validity of KSE formalism relies on a fact that the spatial and temporal coherence of the quantum fluid is primarily dominated by the phase correlator.
As usual, the contribution from density fluctuations turns out to be negligible for the long distance decay of the coherence.
Therefore, the behavior of the  first order coherence function $g^{(1)}(x,t)$ is well approximated by
\begin{eqnarray}    \label{g1_rt2}
g^{(1)} (x,t) = \langle e^{-i \theta (x,t) } e^{i \theta (0,0)}  \rangle
= \langle e^{-i \Delta \theta (x,t)} \rangle ,
\end{eqnarray} 
where $\Delta \theta (x,t) \equiv \theta (x,t) - \theta (0,0)$.
If the phase fluctuation $dW_{\theta}$ behaves as a white noise and has a Gaussian distribution, then the average in Eq.~\eqref{g1_rt2} can    be determined by a standard cumulant expansion up to the second order, which yields
\begin{eqnarray}    \label{g1_rt3}
g^{(1)} (x,t) 
%&=& e^{- {1 \over 2} [ \langle \Delta \theta^2 (x,t) \rangle - \langle \Delta \theta (0,0) \rangle^2 ]}
%\nonumber \\
&\propto& e^{- {1 \over 2} \langle \Delta \theta^2 (x,t) \rangle}
 = e^{- {1 \over 2} C (x,t)} ,
\end{eqnarray}
where a constant factor has been omitted.

%% Numerical methods

To reveal the KPZ scaling of the temporal coherence for the exciton-polariton system, we have performed numerical studies on both GGPE and KSE in one spatial dimension.
The GGPE \eqref{ggpe} is numerically solved using the splitting-flip method: the wave function evolves alternatively as $\psi_k (t) \rightarrow e^{-i T \Delta t} \psi_k (t)$ in the Fourier space and $\psi (x,t) \rightarrow e^{-i V \Delta t} \psi (x,t)$ in the real space, where $T = \hbar^2 k^2 / (2m)$ and $V = g |\psi|^2 +i [ P_0 / (1+|\psi|^2/n_s) - \gamma ]$.
The two evolutions are connected by a Fourier transform, and the noise term is added every time when the real space wave function $\psi (x,t)$ is updated.
The KSE \eqref{rkse} is numerically simulated with the Euler one time step method, $\tilde{\theta} (\tilde{t} + \Delta \tilde{t}) = \tilde{\theta} (\tilde{t}) + \Delta \tilde{\theta}$, in which the nonlinear term is computed using the pseudospectral discretization approach.\cite{Giada2002}
Since the simulation is performed on Eq.~\eqref{rkse} in a dimensionless form, the real units of phase correlation Eq.~\eqref{g1_rt3} are retrieved by employing  the rescaling formula Eq.~\eqref{rs}.
In both GGPE and KSE cases, we implement the simulations on a one-dimensional system of 128 sites with periodic boundary conditions.
To measure the correlation function of the GGPE under dynamic balance, we start from the initial configuration with uniform density and random local phase.
After about $10^5 \sim 10^6$ iterations with time step $\Delta t$=0.001, the system is stabilized at a steady state.

Then the ensemble averaging of temporal correlation $g^{(1)}(x,t)$ is performed over about 1000 sequences of samples of $\psi (x,t)$.
Each sequence of time evolution has a duration of $t$=1000.
For the KSE, we start with a flat distribution of $\theta (x,t)$.
We let the phase field evolves freely at the same step as GGPE up to $t$=1000, during which the phase correlation is measured.
This process is repeated  about 1000 times to get the ensemble average.

%% Numerical results

\section{Temporal correlation and dynamic scaling \label{sec:cor}}

\subsection{Correlation time in Fourier space}

The spatial coherence of non-equilibrium condensates has been explored in an earlier work by us by means of the GGPE and the KSE,\cite{Gladilin2013} where the long distance decay of the spatial coherence was found to be exponential.
It was found that the nonlinear term in the noisy KSE \eqref{kse} can affect the coherence length but it does not change the nature of the decay.
In the present work, we shall concentrate on the temporal correlations.
In analogy with the spatial coherence, we expect the effects of the nonlinearity in the phase equation \eqref{kse} to be more prominent for weak interactions $\tilde \mu \rightarrow 0$.

%%%%%%%%%%%%%%%%%%%%%%%%%%%%%%%%%%%%%%%%%%%%%%%%%
\begin{figure}[htbp]
\centering
\includegraphics[width=0.45\textwidth]{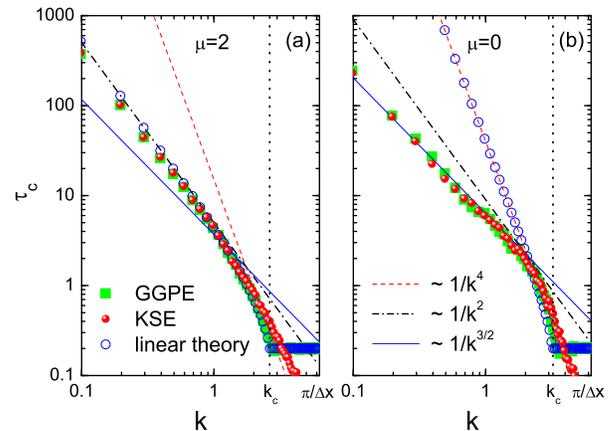}
\caption{(Color online) Characteristic momentum-dependence of correlation time $\tau_c$ in (a) linear regime of $\mu$=2 and (b) nonlinear regime of $\mu$=0.
Three typical power-law dependences of $k$ are plotted by red dashed, black dash-doted and blue solid lines as guide to eyes.
The green squares and red balls are simulations on GGPE and KSE, respectively, and the blue hollow circles denote Bogoliubov theory.
$k_c$ is the bifurcation momentum, and $\Delta x = 0.5$ the discretisation length in real space.}
\label{fig:kpz}
\end{figure} 
%%%%%%%%%%%%%%%%%%%%%%%%%%%%%%%%%%%%%%%%%%%%%%%%%%

As seen in Eq.~\eqref{g1_kt2}, the Bogoliubov theory claims an exponential decay for the temporal correlation.
We therefore introduce a correlation time $\tau_c$ to characterize the time dependence of correlations, and estimate this time scale by fitting the temporal correlation function to an exponential,
\begin{eqnarray}    \label{fit}
g_k^{(1)} (t)=e^{-t/\tau_c (k)}  .
\end{eqnarray}
Fig.~\ref{fig:kpz} displays our numerical results for $\tau_c$ as a function of momentum $k$ in two representative cases, (a) the `linear' regime with $\mu$=2 and (b) the `nonlinear' regime with $\tilde \mu$=0.
The simulations on GGPE and KSE are plotted by the green squares and red balls, respectively, which agree with each other very well in both panels justifying that the spatial and temporal coherences are indeed dominated by the phase fluctuations.
For comparison, the blue open circles represent the Bogolubov theory predictions from Eq.~\eqref{g1_kt2}.
The red dashed, black dash-dotted, and blue solid lines indicate, respectively, the $1/k^4$, $1/k^2$, and $1/k^{3/2}$ relations in the double logarithmic scale.

According to the Bogoliubov theory Eq.~\eqref{g1_kt2}, the long time properties of $g_k^{(1)}$ are classified into two categories depending on the value of $k$.
When $k  > k_c$ ($k_c$ is marked by black dotted lines in Fig.~\ref{fig:kpz}), the profile of $g_k^{(1)}$ shows a time-dependence of $e^{-\Gamma t}$, which corresponds to a trivial constant $\tau_c=1/\Gamma$.
When $k \leq k_c$, the long time behavior turns out to be $e^{-(\Gamma - \omega_k) t}$, and one gets
\begin{eqnarray}    \label{tauc}
\tau_c \sim 1/ \left[ \Gamma - \sqrt{\Gamma^2 - \epsilon_k (\epsilon_k + 2 \mu)} \right] .
\end{eqnarray}
If $\mu \neq 0$, in the region $k \rightarrow 0$, the leading order terms of Eq.~\eqref{tauc} gives $\tau_c \sim \Gamma/(\epsilon_k \mu) \sim 1/k^2$.
This behavior is illustrated in panel~(a) by the gathering of three different symbols along the black dash-dotted line for small momenta, indicating that Bogoliubov theory works well if the interaction energy $\mu$ is sufficiently strong (linear regime for the phase equation).
At the smallest wave vectors however, a small discrepancy between the Bogoliubov theory and the numerics appears. This is expected, because the nonlinear KPZ scaling behavior should take over at large distance and time scales. 

When the interaction energy vanishes ($\tilde \mu \rightarrow 0$) deviations from the linear theory become pronounced at much smaller scales (larger momenta). In Fig. \ref{fig:kpz}, we show Eq.~\eqref{tauc}, giving $\tau_c \sim 2 \Gamma/\epsilon_k^2 \sim 1/k^4$, by the red dashed line and blue hollow circles in panel~(b).
The numerical simulations on GGPE and KSE reveal a marked deviation from the Bogoliubov theory,  immediately below the bifurcation wave vector $k_c$.
Instead of the $\tau_c \sim 1/k^4$ relation, the simulations show unambiguously a $1/k^{3/2}$ dependence, a well-known feature due to the dynamic scaling behavior of KPZ universality class.
In the simulation of GGPE in Fig.~\ref{fig:kpz}, the noise strength is fixed at $D$=0.01.
We have also tested some different values of $D$. Except for a shift in $\tau_c$, we did not find any difference in the scaling.

%%%%%%%%%%%%%%%%%%%%%%%%%%%%%%%%%%%%%%%%%%%%%%%%%
\begin{figure}[htbp]
\centering
\includegraphics[width=0.45\textwidth]{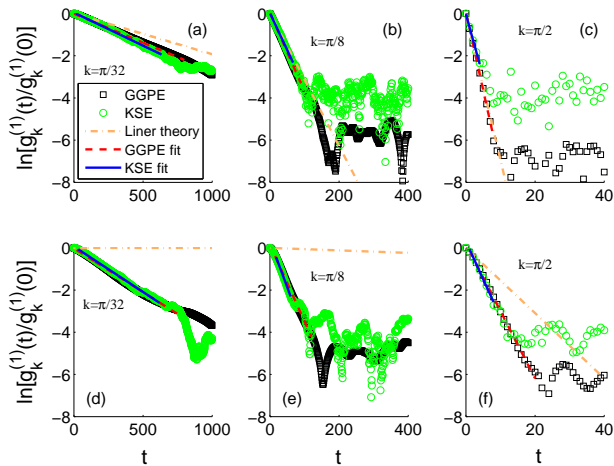}
\caption{(Color online) Extraction of correlation time $\tau_c$ by fitting the temporal correlations in terms of Eq.~\eqref{fit}.
The upper panels (a)-(c) show the correlations for three different values of $k$ when $\mu$=2.
The lower ones (d)-(f) are for $\mu$=0.
The black squares and green circles are from GGPE and KSE simulations, respectively.
Numerical fittings to exponential decays are illustrated by the red dashed and blue solid lines.
The orange dash-dotted lines represent results from Bogoliubov theory as a reference.}
\label{fig:fit}
\end{figure} 
%%%%%%%%%%%%%%%%%%%%%%%%%%%%%%%%%%%%%%%%%%%%%%%%%%

In the nonlinear regime, where the Bogoliubov theory breaks down, it is {\em a priori} no longer guaranteed that the exponential decay \eqref{fit} remains accurate.
In order to address this concern, we present our raw data of simulations together with fitting curves in Fig.~\ref{fig:fit}.
Here we show the time evolutions of function $\ln \left[ g^{(1)}_k (t) / g^{(1)}_k (0) \right]$ for three different momenta, $k=\pi/32$ (the smallest non-zero $k$ in our simulation), $\pi/8$ and $\pi/2$.
The black squares and green circles denote the simulation data of GGPE and KSE, and their fitting results are plotted by the red dashed and blue solid lines, respectively.
One can see in both linear (upper panels) and nonlinear (lower panels) regimes, the straight lines fit quite well on the simulation results up to a cut-off point, after which numerical error spoils the data.
Thus we conclude that even in the nonlinear regime, the temporal correlation decays within our numerical uncertainty like an exponential, and the $1/k^{3/2}$ dependence of $\tau_c$ does characterize the dynamic coherence of the nonequilibrium quantum fluid.

%%

%%%%%%%%%%%%%%%%%%%%%%%%%%%%%%%%%%%%%%%%%%%%%%%%%
\begin{figure}[htbp]
\centering
\includegraphics[width=0.45\textwidth]{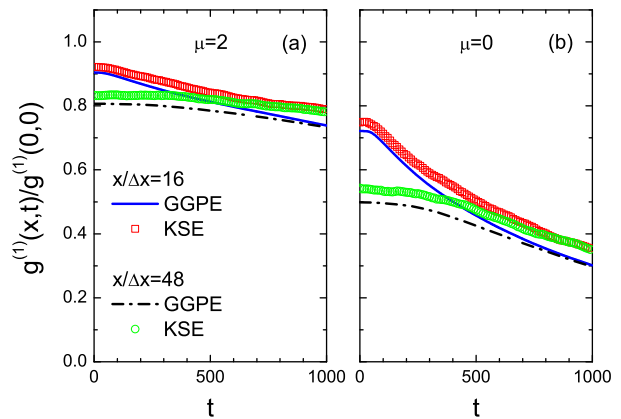}
\caption{(Color online) Time evolution of representative correlation functions in (a) linear regime with $\mu$=2 and (b) nonlinear regime with $\mu$=0.
The blue solid lines depict the correlations at short distance, $x/ \Delta x$=16, while the black dash-dotted lines present those at long distance, $x/\Delta x$=48, both of which are obtained via GGPE.
The corresponding KSE simulations are plotted by the red squares and green circles, respectively.}
\label{fig:g1}
\end{figure} 
%%%%%%%%%%%%%%%%%%%%%%%%%%%%%%%%%%%%%%%%%%%%%%%%%%

\subsection{Scaling function in real space}

So far, we have investigated the scaling of the temporal coherence in momentum space.
While experimentally accessible, it may be hard to make a precise measurement of the coherence time close to zero momentum, where the intensity is very high.
Alternatively, also coherence at fixed distance can be studied.

In Fig.~\ref{fig:g1}, we study the time evolution of spatial correlation in (a) linear regime with $\mu$=2, and (b) nonlinear regime with $\mu$=0.
The blue curves (red squares) display the time-dependence of the correlation for closely spaced points with $x/ \Delta x$=16, obtained with the GGPE (KSE).
For comparison, the correlation between two distant points of $x/\Delta x$=48 is shown by the black dash-dotted curve (green circles) computed with GGPE (KSE) over the same time range.

As one can see, with elapsing time difference, the temporal correlations at short and long spatial distance tend to the same asymptote, though they are very different at equal time.
Here one also notices some small difference between GGPE and KSE results, which we attribute to density fluctuations.

%%%%%%%%%%%%%%%%%%%%%%%%%%%%%%%%%%%%%%%%%%%%%%%%%
\begin{figure}[htbp]
\centering
\includegraphics[width=0.45\textwidth]{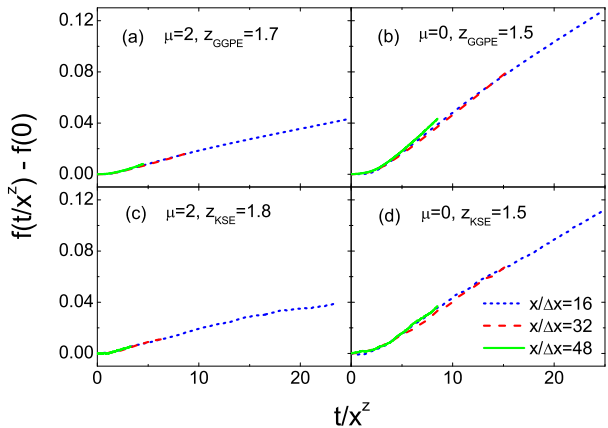}
\caption{(Color online) Dynamic scaling behaviors of GGPE (upper panels) and KSE (lower panels) for different interaction energy $\mu$.
Here $f(t/x^z)$ is the dynamic scaling function, and $z$ is the scaling exponent extracted from numerical simulations (see text).
The three different curves in each panel denote short, intermediate and long distances.
\label{fig:z}
}
\end{figure} 
%%%%%%%%%%%%%%%%%%%%%%%%%%%%%%%%%%%%%%%%%%%%%%%%%%

As discussed in the introduction, the KPZ universality class is characterized by a distinctive scaling function.
With Eq.~\eqref{g1_rt3}, we can connect the scaling function $f$ to the spatial coherence as
\begin{eqnarray}    \label{fkpz}
f(t/x^z) = C(x,t) x^{-2 \chi} = -2 \ln \left[ g^{(1)}(x,t) \right] x^{-2 \chi} .
\end{eqnarray}
Fig.~\ref{fig:z} plots our numerical results of the shifted scaling function $f(t/x^z) - f(0)$ obtained from numerical simulations over GGPE (upper panels) and KSE (lower panels), respectively.
Here we select several separations in real space $x/\Delta x$=16, 32, 48 on the $N$=128 chain with periodic boundary condition.
The three curves can be made to collapse by a proper spatiotemporal rescaling. Specifically, when $\mu$=2, we find $z$ to be 1.7 for the GGPE simulation in panel (a), and 1.8 for the KSE data in panel (c). 
These values are between the Bogoliubov prediction ($z=2$) and the KPZ theory ($z=1.5$). This shows that for our simulations, the finite size effects are still too large to evidence the KPZ scaling. 
On the other hand, when interactions vanish, $\mu$=0, we find the typical KPZ value of $z$=1.5 for both GGPE and KSE. This result proves that the KPZ physics dominates at all length scales over the Bogoliubov physics in the absence of interactions.
In finite size non-equilibrium quantum fluids, it is thus easier to evidence the characteristic KPZ scaling when the interaction energy is small.

\section{Conclusions \label{sec:con}}

To summarise, we have shown that the coherence function of a 1D non-equilibrium quantum fluid is subject to the KPZ dynamic scaling.
The KPZ scaling feature is a consequence of the nonlinearity inherent in the non-equilibrium system. 
It turns out to be a leading effect in the weak interaction regime, dominating the decay of phase correlation across space and time.
In the nonlinear regime, consistent numerical results on scaling exponents have been obtained from the GGPE and KSE.
Especially at weak interaction, the characteristic KPZ scaling behavior of 1+1 dimensions is recovered with $\chi={1 \over 2}$ and $z={3 \over 2}$ in systems of moderate size.
With the increase of interaction strength, the nonlinear effect is, at intermediate distances, overtaken by the linear one, thus validating the Bogoliubov approach.

Our results show that the KPZ universality class is important to describe the coherence properties of 1D non-equlibrium quantum fluids. It therefore looks promising to further investigate the effect of the KPZ physics on the two-dimensional clean \cite{altman} and disordered \cite{janot} systems.

\end{document}